\documentclass{iopjournal}
\usepackage{ulem}

\begin{document}


\title{Probabilistic Deep Learning Framework for Phase Transformation Forecasting aided by In Situ High--\hspace{-1mm}Temperature Microscopy}
\author{Ioannis Kouroudis$^{1*\dagger}$\orcid{0000-0001-9759-6696}, Niki Balestrieri$^{1,2*\dagger}$\orcid{0009-0005-3614-0931}, Stefan Rotzsche$^{2}$, Niccolò Radice$^1$, Peter Mayr$^{2}$, Alessio Gagliardi$^{1*}$}

\affil{$^1$Chair of Simulation of Nanosystems for Energy Conversion, TUM School of Computation, Information and Technology, Technical University of Munich, Munich, Germany}

\affil{$^2$Chair of Materials Engineering of Additive Manufacturing, TUM School of Engineering and Design, Technical University of Munich, Munich, Germany}

\affil{$^*$Author to whom any correspondence should be addressed.}

\affil{$\dagger$These authors contributed equally to this work.}

\email{ioannis.kouroudis@tum.de} 
\email{niki.balestrieri@tum.de} 
\email{alessio.gagliardi@tum.de} 

\begin{abstract}

The mechanical performance of structural materials is governed by their microstructure, which is itself set by the thermal history imposed during processing. Predicting how that microstructure evolves along an arbitrary thermal trajectory, however, remains a challenging problem.
Conventional continuous-cooling-transformation diagrams, for instance, provide only a static description, and they do not exist for all materials, especially emerging and novel ones. Deep-learning approaches address this issue, but they treat image-based characterization, temporal prediction, and microstructure forecasting as separate tasks. Here, we introduce a probabilistic deep-learning framework that unifies these tasks for in situ high-temperature confocal laser scanning microscopy data. The initial surface image is compressed with a $\beta$-variational autoencoder into a low-dimensional latent representation, which, together with the cooling rate and the full temperature history, is passed to a temporal fusion transformer that produces a quantile forecast of the Widmanstätten ferrite ratio. In parallel, a Gaussian process predicts the end-state Widmanstätten ferrite ratio and corrects the forecast equilibrium level, yielding probabilistic prediction intervals. We demonstrate the efficacy of our method on Widmanstätten ferrite evolution in S235 steel across five cooling regimes and showcase significant accuracy. This verifies the relevance of our framework and paves the way for a streamlined and highly accelerated investigation of material evolution under thermal treatment. 
\end{abstract}


\section{Introduction}
For many  materials, the link between processing and performance runs through the microstructure. The phases and morphologies inherited from manufacturing and post-process heat treatments determine strength, toughness, and failure behavior \cite{bhadeshia2017steels,2017materials} as well optical and reactive properties \cite{harth2023optoelectronic,Gryc2026}. In structural materials, and in particular in steels, such as S235, the transformation of austenite during cooling produces a heterogeneous multiphase system, including ferrite and cementite, as well as microconstituents such as pearlite, bainite, and martensite \cite{wang2025high}. In processes that involve complex thermal cycles, such as welding or additive manufacturing (AM), additional structures in these constituents can appear, such as Widmanstätten structure or acicular ferrite. Another factor to consider is the layer by layer deposition in AM, where complex thermal cycles can contain remelting and high cooling rates \cite{adomako2023predicting}. The volume fraction of those phases and microconstituents governs the resulting mechanical properties of the alloy: strength, toughness, and ductility.
An accurate quantification and prediction of these phases and their transformations is therefore fundamental in both industrial applications and research, especially under controlled thermal conditions.
Conventional approaches rely on Continuous Cooling Transformation (CCT) diagrams and macroscopic thermodynamic or kinetic models, such as Johnson Mehl Avrami Kolmogorov formulations and phase-field simulations \cite{karl2025computational, laskowski2021phase}. While well established, these methods suffer from several limitations. A CCT diagram is measured for a single composition and austenitizing condition and assumes continuous cooling at a nominally constant rate; it therefore cannot be transferred to the arbitrary, non-monotonic thermal cycles encountered in welding or AM. Moreover, it reports only global transformation start and finish temperatures and average final phase fractions, typically inferred from dilatometric or thermal signals, and thus carries no information on the morphology, spatial distribution, or grain heterogeneity of the evolving microstructure as it develops experimentally \cite{harrison1989application}. Finally, the considerable experimental effort required to construct such diagrams means that they simply do not exist for many emerging alloys.
High-temperature confocal laser scanning microscopy (HT-CLSM) offers a unique pathway for observing this evolution in situ and in real time \cite{mu2018high}. Yet the extraction of quantitative information and the subsequent forecasting of transformation trajectories under arbitrary thermal histories remain open methodological challenges that the present work seeks to address.

Over the past decade, deep learning has established itself as a powerful tool for automating the analysis of materials' microstructures and predicting specific phase transformation parameters. Nevertheless, the integration with HT-CLSM remains largely unexplored, with most approaches limited to analyzing post-processed samples. In fact, the focus has mostly been on predicting final states rather than forecasting dynamic microstructural evolution during applied thermal cycles. Access to the full transformation trajectory, however, is considerably more informative than the end state alone. It exposes the kinetics of the transformation: onset, growth rate, and saturation, which govern the microconstituent morphologies and hence the final mechanical properties; it allows the response to previously unseen thermal histories to be predicted, rather than interpolated between a few fixed processing routes; and it opens the possibility of monitoring and adapting the thermal cycle while the transformation is still in progress, rather than assessing the outcome only after processing is complete. In the following analysis, we review these developments, grouped by the tasks they address, and identify the gap that the present framework fills.

Recent years have seen increasing use of deep learning to model the relationships between processing, microstructure, and properties in materials, particularly where conventional image analysis and physics-based approaches become difficult to apply efficiently \cite{mayr2022machine, Gryc2026}. One important direction concerns the characterization of microstructures and properties from images, which has developed along three main lines. The first is the segmentation and identification of phases and constituents. Convolutional neural networks have been applied to phase and constituent segmentation in complex-phase steels, with U-Net architectures trained on approximately 50 micrographs annotated with electron backscatter diffraction (EBSD) data, achieving high accuracy in lath-bainite identification \cite{durmaz-2021, munoz-rodenas-2024}, and comparative studies of fully convolutional architectures report similarly accurate segmentation of ferrite and martensite dual-phase steels \cite{ajioka2020development}, while EBSD labeled training data enable high-throughput classification and quantification of more challenging dual-phase microstructures \cite{shen2021generic}. Hybrid and self-supervised strategies have since reduced the dependence on costly scanning electron microscopy (SEM) annotations: a Vision Transformer pre-trained on unlabeled optical micrographs and combined with a convolutional decoder for phase segmentation reaches accuracies comparable to baselines fully supervised on SEM annotations, while using only a sparse set of aligned SEM labels for fine-tuning \cite{ZHANG2024114532}. These approaches, however, assign each pixel to one of only two classes, separating a single constituent of interest from the surrounding matrix. Beyond binary segmentation, deeper convolutional classifiers assign each pixel or region of a micrograph to one of the full set of constituents typically encountered in heat-treated steels: ferrite, pearlite, cementite, austenite, bainite, and martensite, and can further distinguish bainitic subclasses such as granular, degenerated upper, upper, and lower bainite, degenerated pearlite, and incomplete transformation products, although the residual confusion between bainite, martensite, and pearlite at fine scales remains a challenge \cite{phase_3, muller-2020,muller-2021, azimi2018advanced}. Object-detection formulations of the same task have also emerged, with YOLO networks localizing individual cementite particles in spheroidized pearlite for automated metallographic evaluation \cite{liu2025cementite}, and generative adversarial networks have automated grain size quantification from SEM backscatter images \cite{anantatamukala2023generative}. The second line addresses the identification of the heat treatment route itself: transfer learning variants based on ResNet50 and GoogLeNet classify low carbon steel micrographs as annealed, quenched, or quenched and tempered with almost perfect accuracy on optical images \cite{classification_heat_treatment}. The third line couples image information with alloy descriptors to regress scalar transformation parameters, such as the martensite start temperature \cite{ma16030932}; inverse formulations of the same idea recover prior processing conditions and composition from the final microstructure morphology of Fe-Cr-Co alloys \cite{FARIZHANDI2022110799}.

Beyond image based approaches, another class of models targets related quantities using only composition and processing variables as inputs. For the martensite start temperature, ensemble regressors trained on large composition databases match or surpass classical empirical formulae and CALPHAD estimates over wide compositional ranges, with reported errors on the order of 18 K \cite{martensite_starting_temp,start_temperature_3}. Similar approaches have been applied to other transformation temperatures, with a neural network model predicting the $\text{A}_{\text{cm}}$ temperature governing cementite precipitation from austenite in low alloy steels more accurately than existing empirical equations \cite{jeon-2022}. Accuracy and interpretability improve further when expert knowledge is embedded in the feature set. One such study augmented the inputs with metallurgical descriptors: the austenitizing temperature, the thermodynamic driving force of the transformation, and atomic properties such as electronegativity and valence-electron counts, then coupled them with explicit feature selection and SHAP-based interpretation, yielding both higher accuracy and a quantitative ranking of the most influential variables \cite{start_temperature_2}. The same logic has been extended to entire transformation diagrams. A hybrid classifier–regressor reconstructs full time-temperature-transformation (TTT) curves of high alloy steels from composition and austenitizing conditions \cite{ml_temp_diagrams_steel}, and similar models cover carbon and low alloy steels \cite{huang2023machine}. CCT diagrams have likewise been constructed directly from data across steel grades \cite{luukkonen2023gradient, geng2020modeling, zinyagin2024low}. Closer to the process itself, strain induced martensite content has been predicted from ongoing force and temperature signals during cryogenic machining \cite{martensite_content_ml}. A related effort addresses the morphology of the transformation products themselves: a stacked classification-regression model predicts both the type of bainite formed and the fractions of polygonal ferrite, granular bainite, acicular ferrite, and lath bainite from composition and rolling parameters. This can be used to reduce the content of Cr, Ni, Mo, and Cu while preserving the mechanical performance of the steel \cite{CAO2025114642}. Hybrid schemes that couple physical models with machine learning follow the same philosophy, for instance predicting the interlamellar spacing of high carbon pearlitic steels and the mechanical properties that derive from it \cite{qiao-2021}. More recently, physics informed formulations that embed physical descriptors and empirical metallurgic knowledge directly into the learning process have improved the robustness and generalizability of both martensite start temperature prediction and CCT diagram construction \cite{wang2025physically, hedstrom2026physics}. A common feature of all these models is that they take composition, and at most a few global processing parameters, as input. The microstructure that actually develops, and the thermal history that produces it, enter only indirectly through the training labels \cite{ml_martensite_reviews}. This is adequate for alloy screening, but it leaves untouched the question of how a given microstructure evolves in time under a specific thermal trajectory, which is precisely the regime addressed by in situ high-temperature microscopy.

It is therefore natural to address phase transformation as a temporal prediction problem. Long Short-Term Memory (LSTM) architectures, recurrent neural networks designed to retain information over extended sequences, have been used to track austenite decomposition during cooling. The networks learn directly from transformation curves extracted from CCT diagrams and reconstruct the time evolution of the individual phase fractions: ferrite, pearlite, bainite, and martensite \cite{kulawik-2021,wrobel-2022}. The same approach extends to processes where cooling and deformation are coupled. In hot stamping, a network based on Gated Recurrent Units (GRU), a streamlined variant of the LSTM cell, takes as input the local thermal and strain history of a material point and returns the time evolution of its phase fractions. The trained network is then used as a constitutive model inside a coupled finite element simulation of the forming process, producing a full map of microstructure and hardness across the formed component \cite{LI2022107134}. Beyond steels, recurrent, convolutional, and attention based sequence models have been benchmarked on forecasting grain size distribution during grain growth \cite{younes2025predicting}.

The works discussed so far address either the spatial or the temporal dimension of microstructure evolution separately: image based methods extract phase information from individual micrographs, while sequence models predict scalar phase fractions over time. A more recent body of work combines the two by compressing each micrograph into a low dimensional latent representation, typically through a convolutional autoencoder, and modeling its evolution as a sequence in this latent space. Autoencoder-LSTM frameworks have been used to accelerate phase simulations of coupled microstructural problems such as Ostwald ripening, with reported speed-ups exceeding five orders of magnitude relative to direct simulation \cite{Gesch2025AcceleratingPS}, and similar autoencoder-RNN combinations have been benchmarked on spinodal decomposition in binary and ternary mixtures \cite{TIWARI2025113518}. Variations on this scheme replace the autoencoder with principal component representations \cite{montesdeoca2021accelerating}, learn the latent dynamics with neural operators, which also enables extrapolation in time beyond the training window \cite{oommen2022learning}, or skip the compression step entirely and let a convolutional recurrent network operate directly on the image sequence \cite{yang2021self}. Transformer-based variants have been applied to polycrystalline grain evolution during solidification, again as surrogates of phase-field calculations \cite{GAO2024109477}. The same approach has also been ported from simulated to experimental data: in situ SEM image sequences acquired during heat treatment of carbon steels have been used to train a recurrent neural network to forecast the future microstructural state from past observations \cite{ZHANG2024100471}. A related use of transformer architectures targets local properties rather than microstructural fields, mapping infrared thermal histories acquired during wire arc additive manufacturing to the local ultimate tensile strength of Inconel 625 components \cite{keshmiri-2025}.

Taken together, these studies show that deep learning has become an established tool for microstructure analysis and phase transformation modeling in steels. At the same time, most existing approaches still treat image-based characterization, temporal prediction, and the forecasting of evolving microstructures as separate tasks. In high-temperature microscopy, however, these aspects are inherently connected, since the microstructure observed at any point in time depends on the preceding thermal history and on the process variables that drive the transformation. Further, most prior approaches must choose between global context architectures, such as transformers, and recurrent architectures with limited effective memory, such as LSTMs.

In this context, in situ HT-CLSM measurements constitute a qualitative step beyond the data on which the approaches reviewed above are built. Instead of post-mortem micrographs or simulated fields, they deliver temporally continuous image sequences of the transforming surface together with the exact thermal history that produced them, thereby capturing the spatial and the temporal dimensions of the transformation simultaneously, in real time, and under controlled, programmable thermal cycles. Data of this richness naturally call for a model able to exploit both dimensions jointly, which is precisely what motivates the machine learning framework proposed here. The novelty of this work lies in unifying image-based characterization, temporal prediction, and microstructure forecasting into a single pipeline for in situ HT-CLSM data. The initial surface image is compressed through a variational autoencoder into a low dimensional latent representation and supplied, together with the thermal history $T(t)$, to a temporal fusion transformer that outputs a probabilistic forecast of the phase fraction trajectory. Encoding the image as a static covariate conditions the forecast on the specific starting microstructure rather than on a few global descriptors, while the quantile output provides prediction intervals that track the kinetic variability of the transformation. The temporal fusion transformer is chosen precisely because it captures both short and long range temporal dependencies in a single model, eliminating the usual trade-off between recurrence and attention-based architectures. The model can be generalized to many different material processes; the thermal evolution of Widmanstätten ferrite in S235 is taken here as a representative example on which we demonstrate the pipeline.


\section{Methods}

\subsection{In-Situ High-Temperature Microscopy}
\begin{figure}[h!]
    \centering
    \includegraphics[width=1\linewidth]{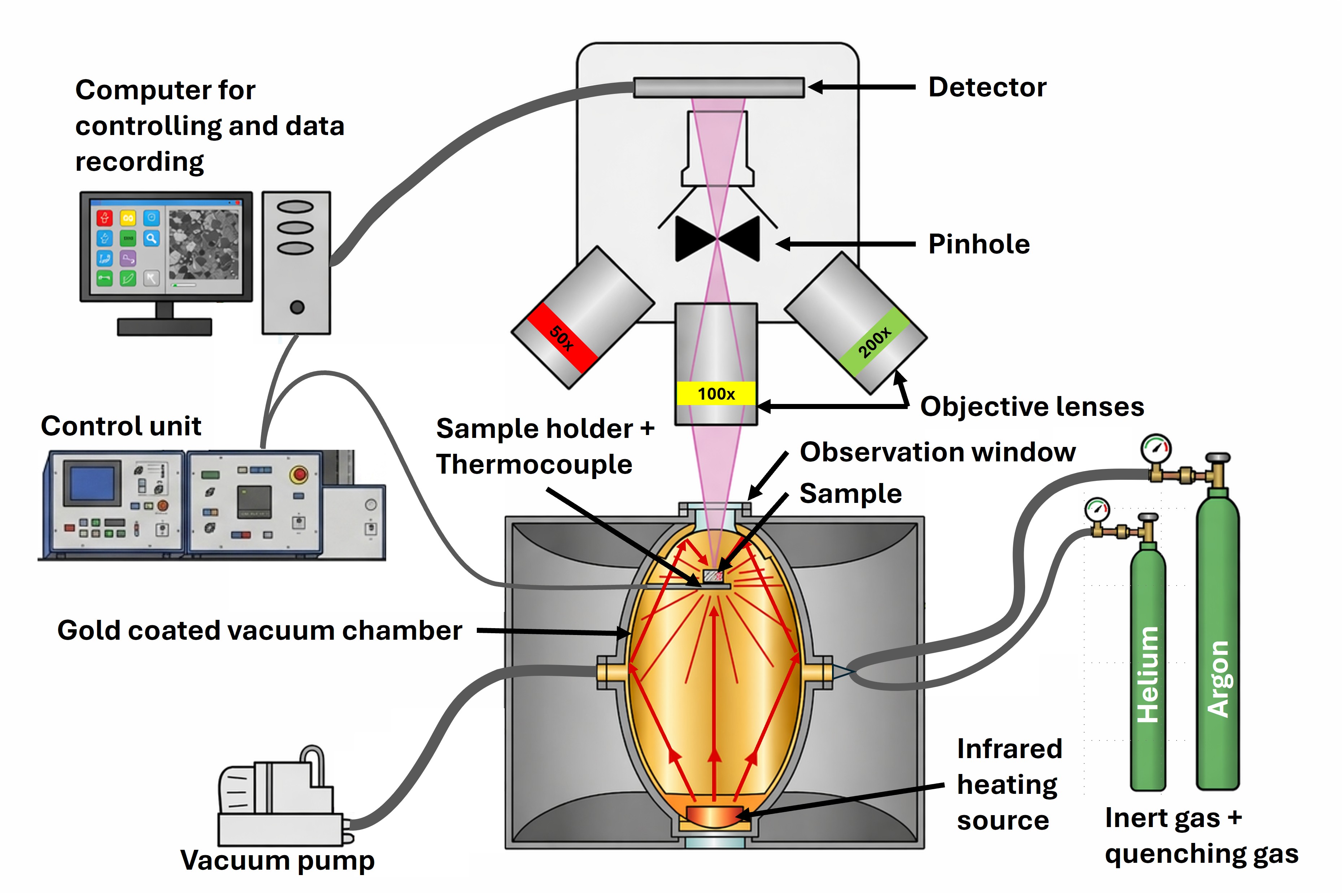}
    \caption{Schematic overview of the high-temperature microscope.}
    \label{fig:HTM}
\end{figure}

The microstructural evolution of steel S235 was examined using a Yonekura SVF16SP HT-CLSM, shown in Figure~\ref{fig:HTM}. The microscope is equipped with a Lasertec VL2000DX confocal scanning laser and a gold coated reflective furnace chamber with an infrared heating source, achieving heating rates up to $100~\mathrm{K\,s^{-1}}$. For quenching or fast cooling, the furnace can be flooded with helium gas to achieve high cooling rates. The sample holder and crucible are made from Al\textsubscript{2}O\textsubscript{3} with a type R thermocouple located beneath the sample holder. Before running an experiment, the chamber was evacuated and flushed with argon (Ar), and experiments were conducted under a continuous Ar flow at positive pressure to maintain an inert gas atmosphere.
Through the observation window, the polished surface of a sample can be observed. When the focal plane is illuminated by a 405 nm laser source, a focused image can be obtained at high temperatures, independent of the sample's thermal radiation. Furthermore, the pinhole suppresses regions that are out of focus. This enables contrast generation, allowing potential phase transformations and other microstructural phenomena at the sample surface to be visualized, and facilitates in situ observation of the sample's microstructure.


\subsection{Experimental Process}
\label{sec:experimental}

For microscopic investigations, the samples were first prepared metallographically. To this end, the samples were cut to a size of $5\times5\times5$~mm, embedded, ground, and polished. To achieve high surface quality, the final polishing step was performed using a 0.06 µm colloidal silica suspension. The samples were then removed from the embedding material and transferred to the furnace chamber. In situ videos were acquired using objectives at magnifications of 100$\times$ and 200$\times$, while a detection optics collected the reflected intensity from the specimen surface. For each experimental run, a continuous video was recorded at 10 frames per second (fps) with a spatial resolution of $1024\times1024$ pixels and stored. The thermocouple data were synchronized with the video frames via timestamps, enabling direct correlation between the temperature history and the microstructural response. A summary of the acquisition parameters is reported in Table~\ref{tab:htm_params}.
\begin{table}[h]
\caption{Experimental and acquisition parameters.}
\label{tab:htm_params}
\centering
\begin{tabular}{ll}
\toprule
\textbf{Category} & \textbf{Parameter} \\
\midrule
Microscopy setup & High-temperature confocal laser scanning microscopy (HT-CLSM) \\
Objectives & Long-working-distance: 100$\times$, 200$\times$ \\
Temperature measurement & Type R thermocouple (beneath sample holder) \\
Atmosphere & Ar \\
Acquisition rate & 10~frames per second (fps) \\
Frame size & $1024\times1024$ pixels \\
Raw file format & AVI \\
\bottomrule
\end{tabular}
\end{table}

\begin{figure}[h]
    \centering
    \includegraphics[width=\linewidth]{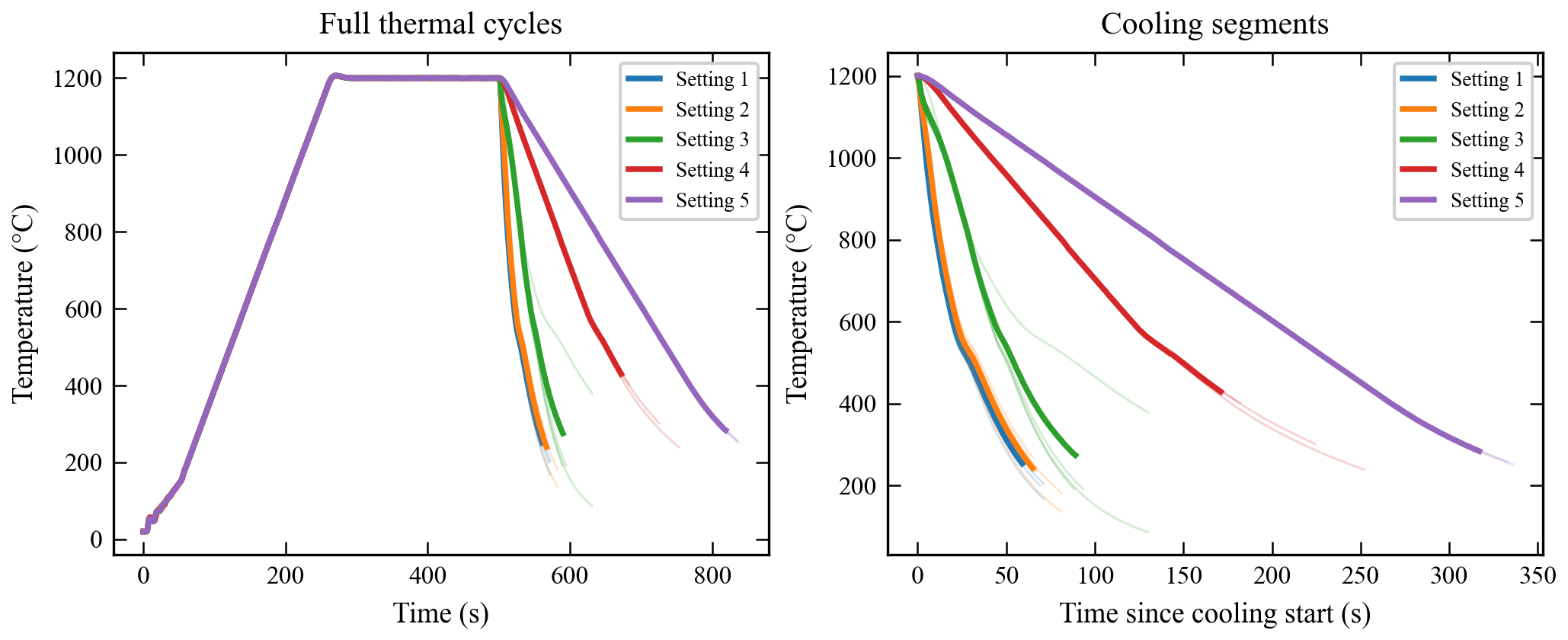}
    \caption{Thermal cycles overview. \textbf{Left:} full temperature histories $T(t)$ for Settings~1--5 (four repeats per setting). \textbf{Right:} cooling segments aligned to the cooling onset. Thick curves show the mean trajectory and faint curves denote individual experiments (S1--S4).}
    \label{fig:thermal_cycles_ab}
\end{figure}

The thermal profiles were designed to be representative of different transformation behaviors and scenarios. All specimens were heated to 1200~$^\circ$C at 20~$\mathrm{K\,s^{-1}}$, held for 4~min, and subsequently cooled to room temperature under controlled conditions. Cooling experiments were grouped into five settings (Settings~1--5), each repeated four times (S1--S4). Figure~\ref{fig:thermal_cycles_ab} summarizes the full temperature histories and the corresponding specific cooling windows.

Across all experiments, the analyzed cooling segment began at $T_{\mathrm{i,c}} = 1200.0 \pm 0.3~^\circ$C, and its onset was highly reproducible, occurring at $t_{\mathrm{i,c}} = 501.40 \pm 0.03$~s (mean $\pm$ standard deviation across all settings). Cooling severity was quantified using the standard $t_{8/5}$ metric, defined as the time required to cool from 800~$^\circ$C to 500~$^\circ$C. The resulting cooling regimes span from fast to very slow conditions, as summarized in Table~\ref{tab:t85}. Settings~1 and~2 yielded comparable fast cooling rates, Setting~3 exhibited intermediate behavior with larger variability, and Settings~4 and~5 produced slow and very slow cooling. Linear regression over the 800--500~$^\circ$C interval confirmed highly consistent cooling in the slow regimes.

\begin{table}[h!]
\caption{Summary of cooling conditions quantified via the $t_{8/5}$ metric across four repeats (S1--S4). Values are reported as mean $\pm$ standard deviation across repeats.}
\label{tab:t85}
\centering
\begin{tabular}{lcccc}
\toprule
\textbf{Setting} & $\boldsymbol{\dot{T}_{8/5}}$ \textbf{(K\,s$^{-1}$)} & $\boldsymbol{t_{8/5}}$ \textbf{(s)} & \textbf{Fit $R^2$} & \textbf{Min/Max $\dot{T}_{8/5}$ (K\,s$^{-1}$)} \\
\midrule
1 & 16.00 $\pm$ 0.90 & 18.80 $\pm$ 1.07 & 0.908 $\pm$ 0.004 & 14.86 / 17.05 \\
2 & 15.25 $\pm$ 1.92 & 19.90 $\pm$ 2.43 & 0.918 $\pm$ 0.006 & 13.61 / 17.65 \\
3 & 11.65 $\pm$ 4.39 & 31.02 $\pm$ 18.67 & 0.965 $\pm$ 0.035 &  5.08 / 14.27 \\
4 &  4.40 $\pm$ 0.10 & 68.34 $\pm$ 1.55 & 0.995 $\pm$ 0.001 &  4.30 /  4.53 \\
5 &  3.02 $\pm$ 0.01 & 99.27 $\pm$ 0.18 & 0.9998 $\pm$ 0.0001 &  3.02 /  3.03 \\
\bottomrule
\end{tabular}
\end{table}
\FloatBarrier

Subsequently, the data were post-processed using a Python and OpenCV-based pipeline. Representative frames were extracted via uniform sampling at 1~fps, yielding approximately 20 frames per experiment. Each frame was converted to grayscale, the contrast normalized via histogram equalization to mitigate radiance variations at elevated temperatures, and cropped to a consistent region of interest of $800\times800$ pixels.

\subsection{Phase Identification and Gradient Masking}
\label{sec:phase_id}
To interpret the model outputs physically, the tracked phase must be identified and quantified from the images. This procedure was performed using an automatic gradient masking routine implemented in Python. As described in Section~\ref{sec:experimental}, HT-CLSM acquisition is continuous over the thermal cycle, with frames sampled at $1$~fps during the cooling segment. In each frame, the routine isolates the Widmanstätten ferrite laths by thresholding the local contrast gradient. Subsequently,  the Widmanstätten ferrite area fraction was computed as the ratio of masked to total pixels (hereafter the Widmanstätten ferrite ratio). The identification of these laths as Widmanstätten ferrite was supported by the synchronized thermocouple record and observations prior to cooling. At high austenitization temperatures around $1200~^\circ$C, Widmanstätten ferrite is prone to form during cooling owing to the large austenite grain size. Applied frame by frame across the cooling segment, this produces a per-frame Widmanstätten ferrite ratio that increases monotonically as the transformation proceeds; this sequence is the ground-truth target used to train and evaluate the model. Throughout this work, the Widmanstätten ferrite ratio is reported either as a fraction in $[0,1]$ or, equivalently, as a percentage.

\begin{figure}[h!]
    \centering
    \includegraphics[width=0.8\linewidth]{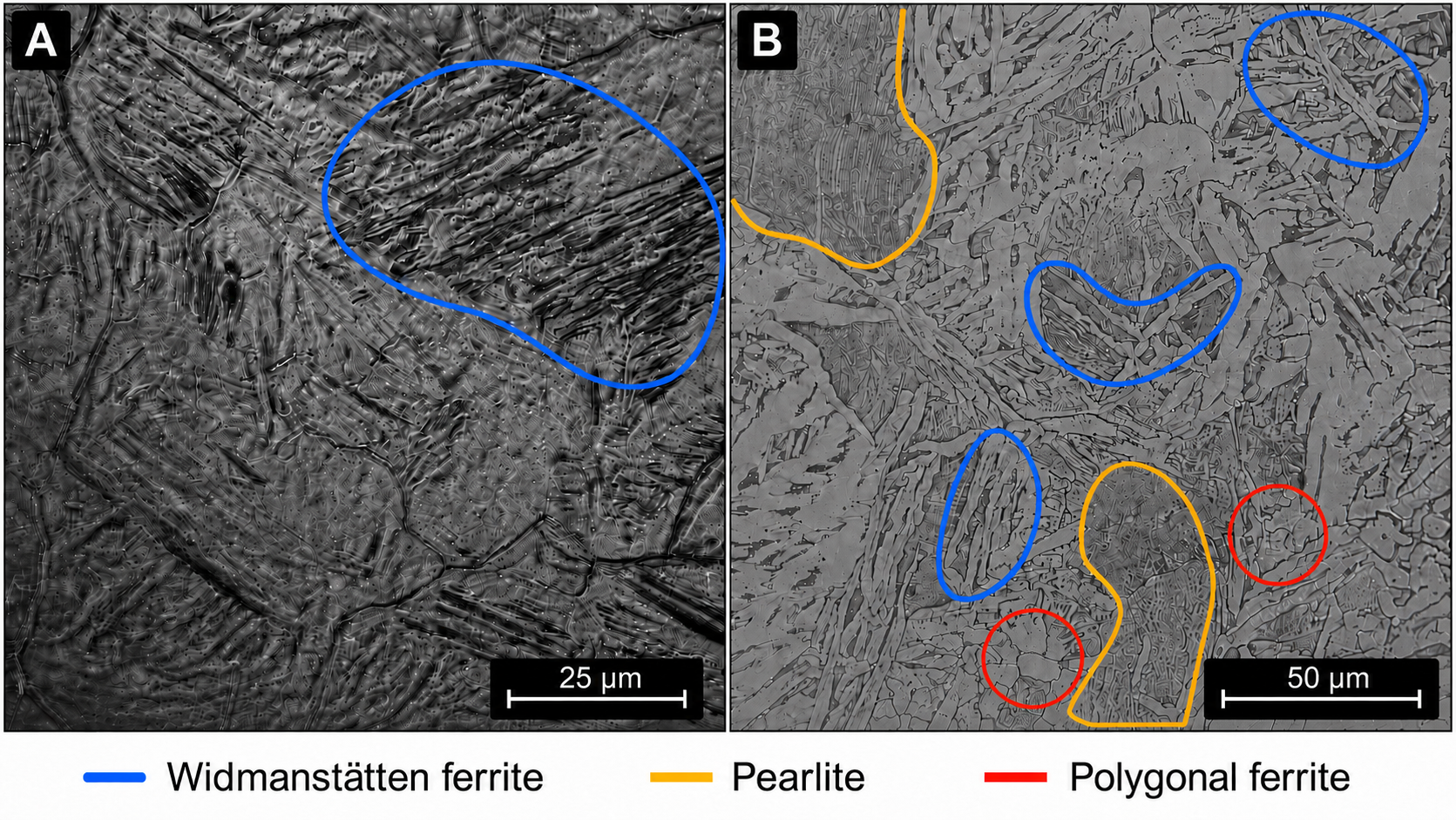}
    \caption{Phase identification by comparison of the in situ HT-CLSM image \textbf{(left)} and the nital-etched optical micrograph \textbf{(right)} for sample S3S1. In the HT-CLSM image, Widmanstätten ferrite appears as laths. The optical micrograph shows the laths of Widmanstätten ferrite (blue circle), pearlite (yellow circle), and polygonal ferrite (red circle).}
    \label{fig:nital}
\end{figure}

The final frame of each experiment, acquired once the sample has cooled to room temperature and the transformation is complete, provides the highest contrast and the least ambiguous segmentation. Therefore, the masked fraction was taken as the empirical end state ratio of Widmanstätten ferrite. To confirm that this measurement is physically meaningful, the final-frame HT-CLSM segmentation was cross-compared with post-mortem optical microscopy of the same sample after nital etching, using a Leica DM6 M. The etched micrograph reveals the full microstructure at room temperature: lamellar pearlite appears dark, while all ferrite morphologies, including Widmanstätten and polygonal ferrite, appear light. The Widmanstätten ferrite is the light lath-shaped constituent, distinguished from the dark lamellar pearlite, and it corresponds directly to the laths segmented in the HT-CLSM image. When the optical measurement is restricted to this constituent, it can be compared directly with the masked HT-CLSM Widmanstätten ferrite fraction (Figure~\ref{fig:nital}).

\subsection{Data-Driven Pipeline}

The pipeline processes two inputs: an image of the initial morphology, and the heating and cooling evolution as a temperature sequence $T(t)$. Because the image is too large and noisy to use directly as a model input, it is compressed using a variational autoencoder (VAE) into a low-dimensional latent representation. Beyond compression, the Gaussian structure of the VAE latent space has a smoothing and clustering effect: similar images are mapped to nearby latent vectors, ensuring cohesion and continuity of the representation. This latent representation, together with the cooling rate, serves as the static (time-invariant) context for the temporal fusion transformer (TFT). To increase the information content, the temperature sequence $T(t)$ is additionally supplied as a dynamic (time-varying) input. The TFT captures local temporal dependencies via a Long Short-Term Memory (LSTM) encoder and global dependencies via multi-head attention, ultimately producing a quantile forecast of the Widmanstätten ferrite ratio sequence.

In parallel, the same latent image representation and the cooling rate are passed to a Gaussian process that predicts the end-state Widmanstätten ferrite ratio as a single scalar. The two branches are then combined: the TFT supplies the dynamic evolution profile, while the Gaussian process corrects the equilibrium level, which the purely dynamic TFT tends to under- or over-estimate. The onset of equilibrium is identified from the gradient of the median TFT trajectory. The transformation is taken to have saturated at the first time step that satisfies two conditions simultaneously: the absolute gradient of the median trajectory falls below a small tolerance, $|\,\mathrm{d}\hat{y}_{0.5}/\mathrm{d}t\,| < \varepsilon$, and the elapsed time exceeds a fixed guard, $t > 20 \,\mathrm{s}$. The temporal guard is required because the gradient is also near zero throughout the flat pre-transformation region; without it, the quiescent interval preceding the transformation would be misidentified as the saturated state. From the detected saturation point onward, the median prediction is replaced with the GP equilibrium estimate.
Because both models are innately stochastic, each prediction carries an uncertainty interval: the TFT quantiles describe the uncertainty of the transient trajectory, while the GP posterior describes the uncertainty of the equilibrium level.
The final output is a Widmanstätten ferrite ratio sequence with corrected point estimates and quantile-based prediction intervals, enabling probabilistic characterization of microstructural evolution under arbitrary thermal histories.

\FloatBarrier
\begin{figure}[h!]
    \centering
    \includegraphics[width=\linewidth]{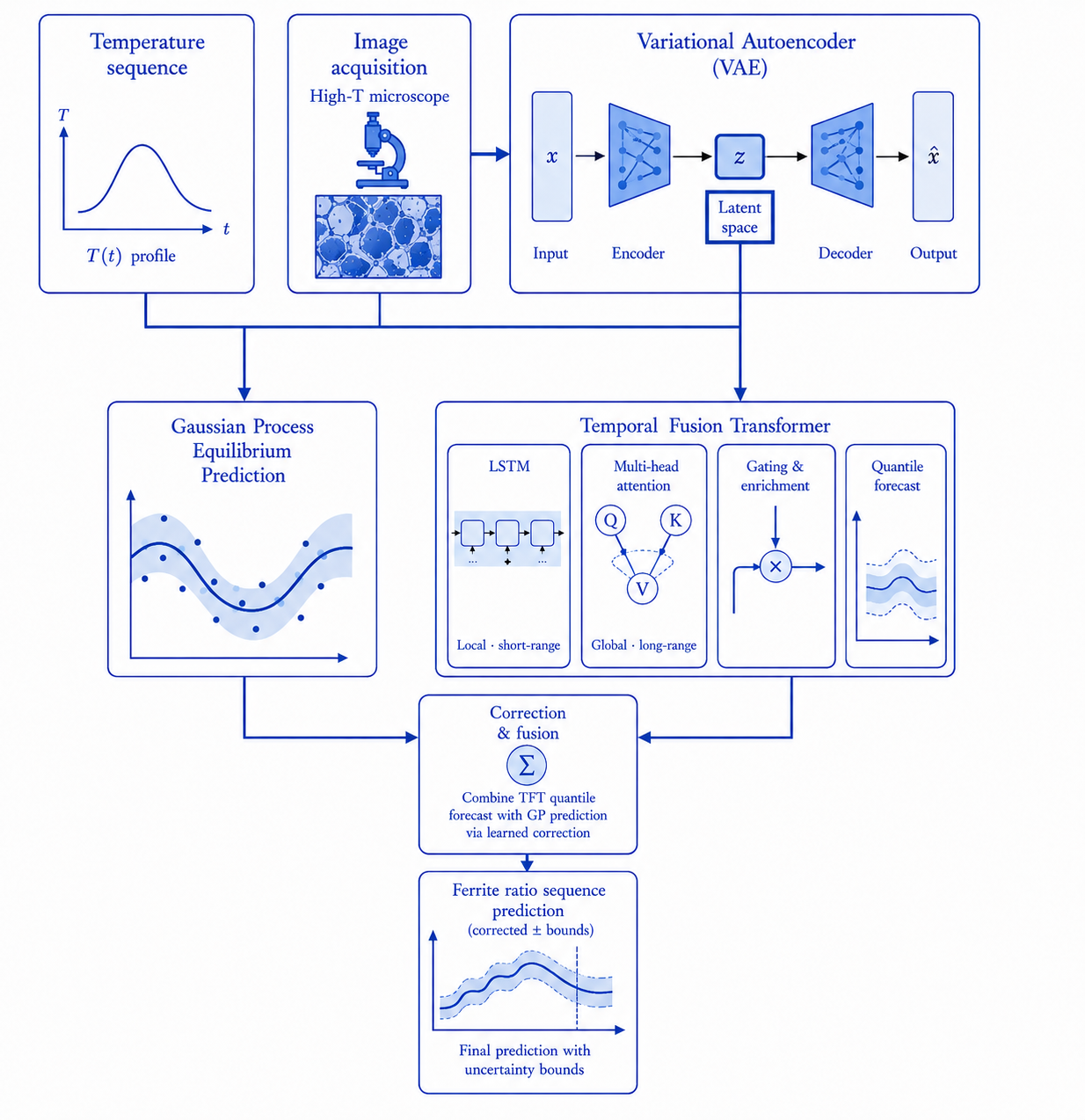}
    \caption{Widmanstätten ferrite evolution prediction pipeline. The initial surface image is compressed with a variational autoencoder (VAE) into a low-dimensional latent representation. This representation, together with the cooling rate and the temperature history $T(t)$, is passed to a temporal fusion transformer (TFT) that produces a quantile forecast of the Widmanstätten ferrite ratio. In parallel, a Gaussian process predicts the end-state (equilibrium) Widmanstätten ferrite ratio from the same latent representation and the cooling rate, and its estimate is fused with the TFT forecast to correct the equilibrium level. The final output is a Widmanstätten ferrite ratio sequence with corrected point estimates and quantile-based prediction intervals.}
    \label{fig:pipeline}
\end{figure}
\FloatBarrier

\subsection{Algorithms Used}

\subsubsection{Variational Autoencoder}
Variational autoencoders (VAEs) are a class of variational dimensionality-reduction models that use convolutional and transposed-convolutional layers to learn a low-dimensional representation of high-dimensional images \cite{kingma2013auto}. The image is encoded into a latent space of distributions and then reconstructed to its original shape. In addition to the reconstruction loss, a Kullback--Leibler (KL) divergence between the approximate posterior $q_{\phi}(\mathbf{z}\mid\mathbf{x})$ and a standard normal prior $p(\mathbf{z})$ smooths the embedding and reduces overfitting. This yields a latent manifold that is both smooth and continuous, so that similar images are mapped to nearby regions, avoiding the abrupt discontinuities characteristic of unregularized embeddings.
In this work, we use a $\beta$-VAE \cite{higgins2017beta, burgess2018understanding}, an extension of the standard VAE in which the KL term is up-weighted by a scalar hyperparameter $\beta$, imposing stronger compression of the posterior toward the prior. In our implementation, the encoder maps each $800\times800$ grayscale frame to a latent space of dimension $d = 20$, chosen to balance reconstruction fidelity and representational compactness, and is trained for 5000 epochs with $\beta = 3$. Consistent with the setting-level cross-validation protocol, the $\beta$-VAE was refit independently within each fold using only the initial frames of the training settings; frames belonging to the held-out setting never entered encoder training. As the encoder is unsupervised, it has no access to the Widmanstätten ferrite ratio targets at any stage, so the latent representation introduces no label leakage between training and test data.

\subsubsection{Gaussian Processes}
The surrogate model for the equilibrium ratio is a Gaussian process (GP), chosen for its robustness against overfitting and the well-established accuracy of global kernel methods. As a data-driven stochastic model, the GP represents predictions as a posterior distribution obtained via Bayes' theorem, with both the prior and the likelihood expressed as Gaussians whose hyperparameters are optimized by maximizing the marginal likelihood over the measured samples. This formulation yields inherent uncertainty quantification, which represents a particularly valuable property when working with small datasets. In this work, the GP takes the VAE latent representation and the cooling rate as inputs and employs a Multi-Layer Perceptron (MLP) kernel \cite{williams1996}, with hyperparameters optimized to maximize the marginal likelihood. 

\subsubsection{Temporal Fusion Transformer}
The temporal fusion transformer (TFT) is a deep learning architecture designed for interpretable multi-horizon time-series forecasting \cite{TTF}.
It integrates recurrent layers, attention mechanisms, and gating structures to capture both long- and short-term interactions. The model processes static and time-varying features through dedicated encoders and uses variable-selection networks to identify the most informative predictors at each stage. A further advantage of the TFT is its probabilistic forecasting framework: instead of producing a single prediction, the model outputs multiple quantiles of the target distribution. This is particularly valuable in physical and industrial processes where measurement noise and process variability are present. The model is trained with the quantile (pinball) loss over the quantile set $\{0.05, 0.25, 0.5, 0.75, 0.95\}$, which directly yields the 25--75\% and 5--95\% prediction intervals reported in \Cref{fig:ferrite_evolution}.

In this study, we used the initial surface microscopy image, suitably reduced, along with the cooling rate as static variables; this allows the model to account for both different cooling regimes and different initial microstructures. As a dynamic covariate, we used the temperature time series. Although this introduces some redundancy with the static cooling rate, the variable-specific processing of static and dynamic covariates extracts complementary information, providing a higher information density for the model.

\FloatBarrier

\section{Results and Discussion}
\subsection{Microstructural Evolution and Phase Fraction Analysis}
\begin{figure}[h!]
 \centering
 \includegraphics[width=1\linewidth]{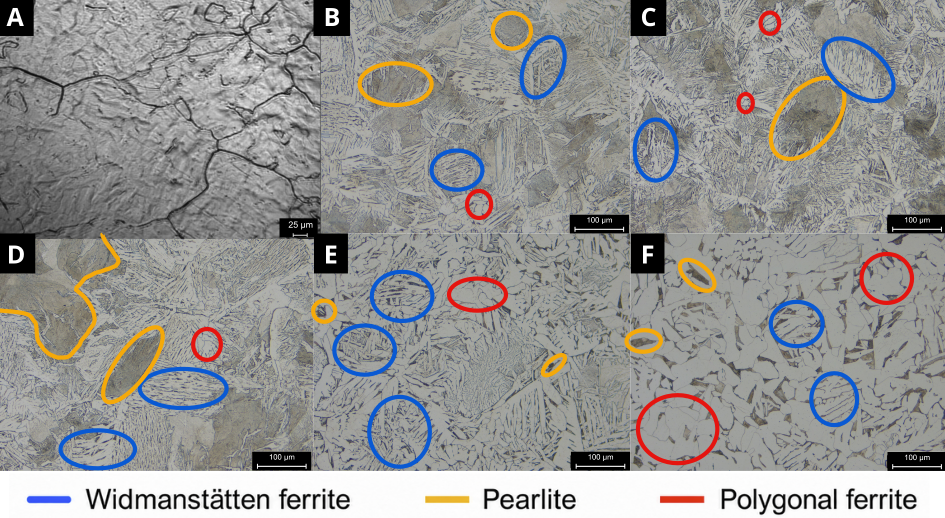}
 \caption{Overview of microstructures from S235 steel. A)~Prior-austenite grain structure at 1200~$^\circ$C, captured in situ by HT-CLSM during the experiment, showing the large austenite grain size. B--F)~Optical micrographs after cooling to room temperature for Settings~1--5, respectively. Blue circle: Widmanstätten ferrite; red circle: polygonal ferrite; yellow circle: pearlite.}
 \label{fig:MS_evolution}
\end{figure}

\begin{figure}[h!]
 \centering
 \includegraphics[width=0.8\linewidth]{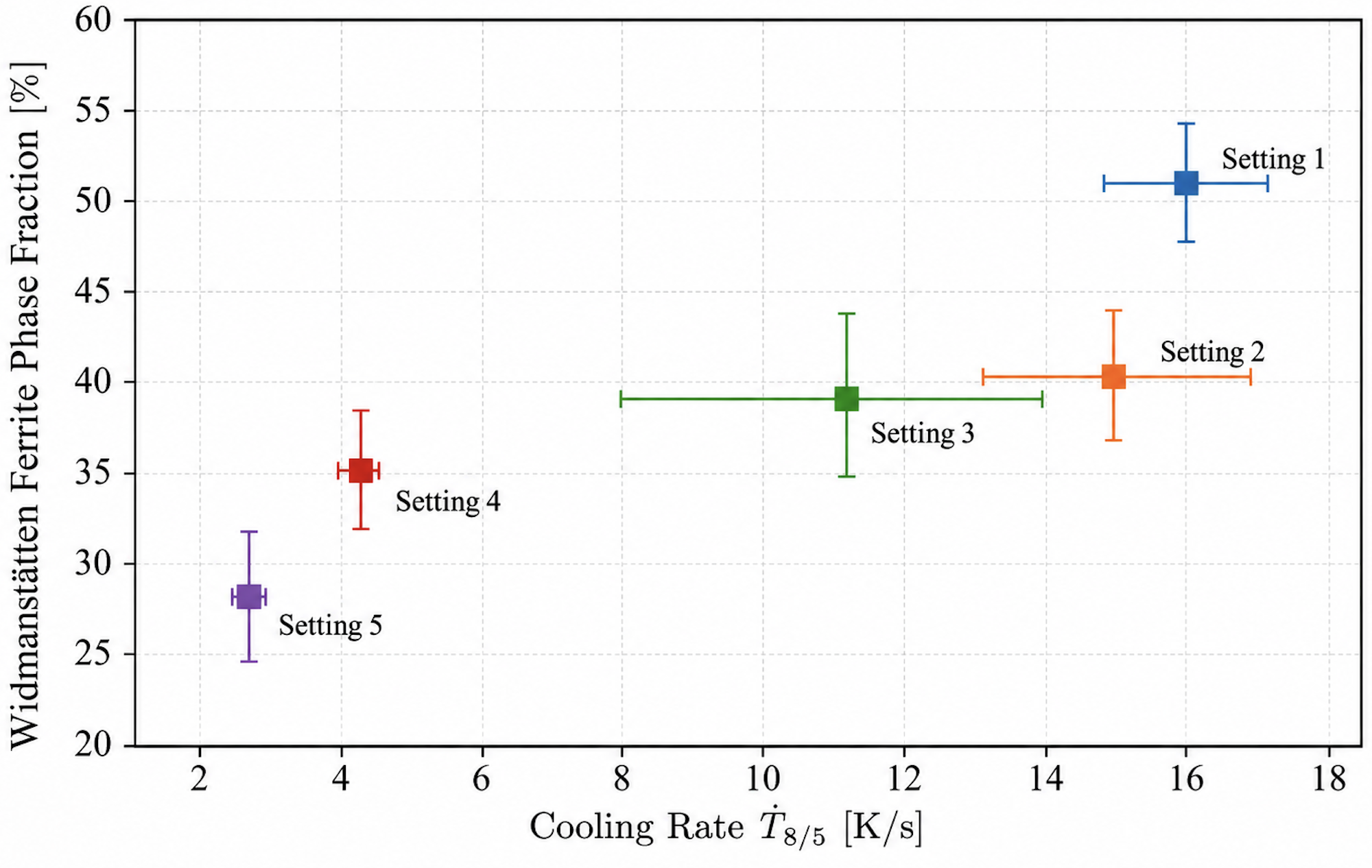}
 \caption{Widmanstätten ferrite phase fraction as a function of the cooling rate $\dot{T}_{8/5}$ for Settings~1--5. Markers denote the mean over the four repeats (S1--S4) of each setting; horizontal error bars indicate the standard deviation of the cooling rate, and vertical error bars the standard deviation of the measured ferrite fraction. The ferrite fraction increases monotonically with cooling rate.}
 \label{fig:WFoverT85}
\end{figure}

\Cref{fig:MS_evolution} summarizes the microstructural evolution of S235 steel after austenitization at 1200~$^\circ$C, followed by thermal cycles with $t_{8/5}$ cooling times ranging from 18.8~s to 99.3~s (Settings~1-5). The coarse grain structure produced during austenitization is shown in \Cref{fig:MS_evolution}A. The elevated austenitization temperature led to pronounced grain coarsening, creating favorable conditions for the formation of Widmanstätten ferrite during subsequent cooling. \Cref{fig:MS_evolution}B-F show the resulting room-temperature microstructures for $t_{8/5}$ times of 18.8~s, 19.9~s, 31.0~s, 68.3~s, and 99.3~s, respectively. Three major constituents were identified: Widmanstätten ferrite, polygonal ferrite, and pearlite.
Quantitative image evaluation revealed a continuous decrease in the Widmanstätten ferrite fraction with increasing $t_{8/5}$ time. The measured Widmanstätten ferrite contents were 51\%, 40\%, 39\%, 35\%, and 28\% for conditions B--F, respectively. Building on this trend, \Cref{fig:WFoverT85} makes the dependence explicit by plotting the measured Widmanstätten ferrite phase fraction against the cooling rate $\dot{T}_{8/5}$ during the $t_{8/5}$ interval: the ferrite fraction increases monotonically with cooling rate. The highest fraction, approximately 51\%, was obtained for Setting~1 at a cooling rate of about $16~\mathrm{K\,s^{-1}}$, while the lowest, approximately 28\%, was measured for Setting~5 at about $3~\mathrm{K\,s^{-1}}$; intermediate thermal conditions yielded fractions between 35\% and 40\%.

This trend is consistent with the transformation behavior of low carbon structural steels. Higher cooling rates suppress diffusional ferrite growth at elevated temperatures and increase the driving force for the formation of ferrite side plates, thereby promoting Widmanstätten ferrite. Conversely, slower cooling rates provide sufficient time for the development of polygonal ferrite and pearlite, reducing the relative fraction of Widmanstätten ferrite. Notably, Widmanstätten ferrite remained present under all investigated conditions; this persistence is attributed to the coarse prior-austenite grain structure generated at 1200~$^\circ$C, since large austenite grains facilitate side-plate growth from grain boundaries and therefore promote Widmanstätten ferrite formation across a broad range of cooling rates. The agreement between the quantified phase fractions and established metallurgical mechanisms confirms that the gradient masking routine captures physically meaningful microstructural variations, providing not only automated phase identification but also quantitative targets that can be related directly to process parameters and thermal histories.

\subsection{Reduced-Order Representation and Forecasting Performance}
\begin{figure}[h!]
 \centering
 \includegraphics[width=\linewidth]{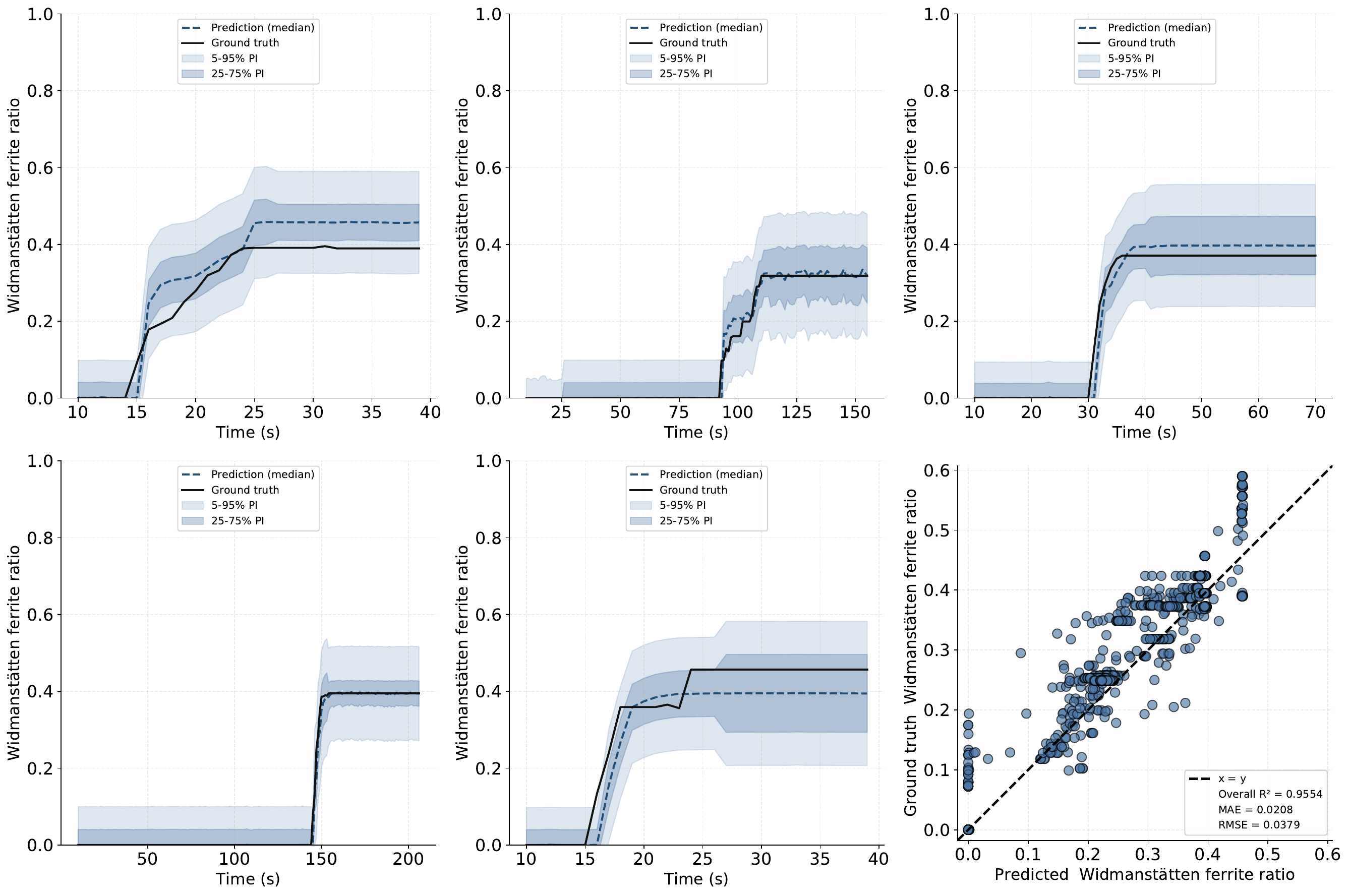}
 \caption{Ground truth and predicted Widmanstätten ferrite ratio over time for a representative evolution from every thermal treatment setting. The solid line is the median prediction; the shaded bands show the 25--75\% and 5--95\% prediction intervals. The bottom-right graph shows the true-versus-predicted Widmanstätten ferrite ratio across all time steps in the test set, aggregated across folds. The dashed line marks perfect agreement ($y = x$).}
 \label{fig:ferrite_evolution}
\end{figure}

Prior to model training, the pre-heating surface images were compressed with a $\beta$-VAE to extract a compact representation of the initial microstructural state. The value of $\beta$ was determined through statistical search, including the no $\beta$ case.
To assess generalizability, a 5-fold cross-validation was employed, with folds defined at the setting level: the four repeats of each setting were kept together, ensuring the model was always evaluated on cooling settings not seen during training. This grouping prevents information leakage between training and test data and simulates real-world deployment on unseen thermal histories.
The predictive performance of the equilibrium-corrected TFT model was evaluated across all test samples by comparing the median predicted Widmanstätten ferrite ratio against the corresponding ground-truth measurements. As shown in the bottom-right panel of \Cref{fig:ferrite_evolution}, the predictions cluster tightly around the identity line over the full range of observed ratios (0 to 0.6), yielding an overall coefficient of determination of $R^2 = 0.955$, a mean absolute error (MAE) of 0.021, and a root-mean-square error (RMSE) of 0.038. The model generalizes well across the intermediate range (0.1--0.4), which constitutes the majority of the dataset. Some scatter is visible at the extremes ($>0.45$ and $<0.1$). Part of this scatter is an expected consequence of the setting-level cross-validation. When an extreme setting (1 or 5) is held out, the Gaussian process must extrapolate the equilibrium level beyond the cooling-rate range seen during training, which is inherently less reliable than interpolation within the intermediate regimes. For fractions above 0.45, the scatter additionally reflects the relative rarity of high-Widmanstätten-ferrite end states and is expected to diminish with additional data. The low-fraction regime ($<0.1$) is intrinsically more challenging, as the Widmanstätten ferrite phase appears abruptly via a displacive transformation, similar to martensite formation, making the onset of the transformation harder to localize in time. This error is nevertheless small, and the bulk of the process of practical interest lies between ratios of 0.1 and 0.4, which are predicted with high accuracy.
\FloatBarrier
To assess the model's ability to capture the dynamics of the transformation along individual thermal histories, \Cref{fig:ferrite_evolution} presents the predicted Widmanstätten ferrite ratio time series for five representative test sequences, one from each cooling setting, spanning different thermal cycles and initial microstructures. In all cases, the median prediction closely tracks the ground truth throughout the full sequence, correctly identifying the transformation onset, the sigmoidal growth regime, and the saturation plateau. The probabilistic intervals further demonstrate the model's value: the 25--75\% interval remains narrow throughout the pre-transformation and plateau regions. It widens during the active transformation phase, where kinetic variability is highest. The broader 5--95\% interval consistently envelops the ground-truth trajectory in all sequences, indicating that, for the sequences shown, the model's uncertainty representation is neither overconfident nor excessively conservative. The asymmetry of the intervals during the transformation ramp suggests that the model encodes the skewed uncertainty inherent to nucleation-driven phase growth.
\subsection{Discussion}
Taken together, the two sets of results are mutually reinforcing. The materials analysis establishes a monotonic increase in the Widmanstätten ferrite fraction with cooling rate, governed by the competition between displacive side-plate growth and the formation of diffusional polygonal ferrite and pearlite in a coarse prior austenite grain structure. The data-driven framework reproduces precisely this behavior, and its two-branch design mirrors the two physical regimes: the Gaussian process, conditioned on the cooling rate and the latent image representation, learns the equilibrium relationship between cooling rate and end-state fraction quantified in \Cref{fig:WFoverT85}, while the temporal fusion transformer captures the transient kinetics of how that end state is approached. The equilibrium correction is therefore not a purely numerical adjustment but reflects a physically meaningful separation between the final phase fraction, largely set by the cooling rate, and the time-resolved path to reach it.
The probabilistic formulation is the principal methodological advance over deterministic CCT-based descriptions and single-point regression models. Conventional CCT diagrams provide a static, population-level summary of transformation behavior, whereas the present framework returns a probabilistic forecast for each thermal trajectory, including its uncertainty. This distinction also sharpens the research gap: existing approaches summarize transformation behavior or provide single-point estimates, but they do not offer trajectory-specific, uncertainty-aware forecasts. The narrow intervals in the plateau regions, where the transformation is effectively complete, and the widening, skewed intervals during the ramp indicate that the model distinguishes deterministic from stochastic phases of the process and approximates the aleatoric uncertainty associated with nucleation. For decision support in process design, this distinction is valuable: it identifies not only the expected microstructural outcome but also the thermal regimes in which that outcome is least certain.
Several limitations should be acknowledged and defined to scope the present proof of concept. The study covers a single steel grade (S235) and a single tracked constituent (Widmanstätten ferrite), and the dataset comprises five cooling settings, each repeated four times, so the extremes of the observed range are sparsely sampled. The ground truth is obtained from a contrast-gradient masking routine applied to the polished surface; while cross-validated against nital-etched optical microscopy, it reflects a two-dimensional surface section rather than the bulk, and is sensitive to the chosen thresholding parameters. The VAE compression also entails a controlled loss of spatial detail, trading reconstruction fidelity for a compact, low-dimensional representation. Generalization across steel grades, additional phases, and more complex (e.g.\ non-monotonic or cyclic) thermal histories remains to be demonstrated and motivates the extensions outlined below.

\FloatBarrier
\FloatBarrier
\section{Conclusion}
In this work, we introduced a multi-model deep learning framework to characterize and predict microstructural evolution in S235 steel under controlled thermal conditions, with a particular focus on high-temperature confocal laser scanning microscopy data. By integrating image-based descriptors with thermal histories, the proposed approach bridges the gap between equilibrium microstructure analysis and dynamic phase-transformation prediction.
The results demonstrate that the framework precisely captures not only the final Widmanstätten ferrite fraction but also the full temporal evolution of the transformation, including the onset and transient regimes. The probabilistic character of the model further enables the quantification of uncertainty, which is essential for real-world applications where experimental variability and noise are unavoidable. Importantly, the use of a reduced-order image representation proved effective for extracting relevant features while maintaining computational efficiency, allowing the model to make predictions conditioned on the specific starting surface rather than reducing a complex, heterogeneous material to a generic set of descriptors.
Future work will focus on extending the framework to additional phase constituents beyond Widmanstätten ferrite and on adding a guiding-optimization functionality that allows the user to tune the thermal profile online to achieve a desired phase mixture. Ultimately, the proposed methodology provides a step toward data-driven decision support in materials processing and offers a scalable pathway to accelerate the design and optimization of heat treatments.


\funding{The authors gratefully acknowledge funding from the following sources:
The Chair of Materials Engineering for Additive Manufacturing provided funding and conducted the experiments.
MENTOR, European Union’s Horizon Europe (HORIZON) Marie Sklodowska-Curie Actions Doctoral Networks (MSCA-DN) HORIZON-MSCA-2023-DN-01 call, under the Grant Agreement number 101169056. The project duration is from 1 October 2024 to 30 September 2028.}

\roles{}
\begin{table}[ht]
\centering
\label{tab:credit}
\begin{tabular}{p{4.5cm}p{10cm}}
\hline
\textbf{Contribution} & \textbf{Author(s)} \\
\hline
Conceptualization & Niki Balestrieri, Ioannis Kouroudis \\
Methodology &  Niki Balestrieri, Ioannis Kouroudis \\
Software &  Niki Balestrieri, Ioannis Kouroudis \\
Validation &  Niki Balestrieri, Ioannis Kouroudis, Stefan Rotzsche, Niccolò Radice\\
Formal analysis &  Niki Balestrieri, Ioannis Kouroudis, Stefan Rotzsche, Niccolò Radice\\
Investigation & Niki Balestrieri, Ioannis Kouroudis, Stefan Rotzsche\\
Resources & Stefan Rotzsche, Alessio Gagliardi, Peter Mayr \\
Data curation & Niki Balestrieri, Ioannis Kouroudis, Stefan Rotzsche, Niccolò Radice \\
Writing -- Original Draft & Niki Balestrieri, Ioannis Kouroudis, Stefan Rotzsche, Niccolò Radice, Alessio Gagliardi \\
Writing -- Review \& Editing & Niki Balestrieri, Ioannis Kouroudis, Stefan Rotzsche, Niccolò Radice, Alessio Gagliardi, Peter Mayr\\
Visualization & Niki Balestrieri, Ioannis Kouroudis, Niccolò Radice \\
Supervision & Ioannis Kouroudis, Stefan Rotzsche, Alessio Gagliardi, Peter Mayr\\
Project administration &  Alessio Gagliardi \\
Funding acquisition &  Alessio Gagliardi, Peter Mayr  \\
\hline
\end{tabular}
\end{table}

\FloatBarrier

\data{Data and code will be made available through GitHub upon acceptance of the publication.}


\bibliographystyle{ieeetr}
\bibliography{bibliography}

\end{document}